\documentclass[%
 reprint,
 amsmath,amssymb,
 aps,
]{revtex4-2}
\usepackage{graphicx}
\usepackage{dcolumn}
\usepackage{bm}

\usepackage{xcolor}
\usepackage{caption}
\usepackage{subcaption}
\usepackage{algpseudocode}
\usepackage{algorithm}
\usepackage[T1]{fontenc}

\begin{document}

\preprint{APS/123-QED}

\title{Synthesizing an Optimal Spin Qubit Shuttling Bus Architecture\\for the Surface Code}

\author{Pau Escofet}
 \email{pau.escofet@upc.edu}
\author{Eduard Alarcón}
\author{Sergi Abadal}
\affiliation{%
 Universitat Politècnica de Catalunya,\\Barcelona, Spain
}%

\author{Andrii Semenov}
\author{Niall Murphy}
\author{Elena Blokhina}
 \altaffiliation[Also at ]{University College Dublin}
\affiliation{%
 Equal1 Labs,\\Dublin, Ireland
}%

\author{Carmen G. Almudéver}
\affiliation{%
 Universitat Politècnica de València,\\València, Spain
}%

\begin{abstract}
As quantum computers scale toward millions of physical qubits, it becomes essential to robustly encode individual logical qubits to ensure fault tolerance under realistic noise. A high-quality foundational encoding allows future compilation techniques and heuristics to build on optimal or near-optimal layouts, improving scalability and error resilience. In this work, we synthesize a one-dimensional shuttling bus architecture for the rotated surface code, leveraging coherent spin-qubit shuttling, following a novel methodology we name \emph{Quantum Reverse Mapping}. We formulate a mixed-integer optimization model that yields optimal solutions with relatively low execution time for small code distances, and propose a scalable heuristic that matches optimal results while maintaining linear computational complexity. We evaluate the synthesized architecture using architectural metrics, such as shuttling distance and cycle time, and full quantum simulations under realistic noise models, showing that the proposed design can sustain logical error rates as low as $2\times 10^{-10}$ per round at code distance 21, showcasing its feasibility for scalable quantum error correction in spin-based quantum processors.
\end{abstract}

\maketitle

\section{Introduction}
\label{sec:introduction}

Quantum error correction (QEC) is widely recognized as a requisite for scaling quantum processors to the fault-tolerant regime  \cite{PhysRevA.52.R2493, error_correction, nielsen2010quantum}. By detecting and correcting errors in quantum states, QEC decreases the error rates and makes the quantum states robust for the large-scale application of quantum algorithms \cite{gidney2021factor, gidney2025factor}.

Among the many families of QEC codes, topological surface codes stand out for their high threshold, reliance on local nearest-neighbor interactions, and the existence of efficient decoding algorithms \cite{bravyi1998quantum, dennis2002topological, fowler2009high, fowler2012surface, litinski2019game, google2023suppressing}.

The performance of an error-corrected logical qubit depends on multiple factors, including the chosen error-correction code, the underlying noise processes, and the available architectural primitives and connectivity. Much of the current state-of-the-art research focuses on how to encode a logical qubit as robustly as possible given a particular code and a fixed physical architecture. For instance, the surface code requires only nearest-neighbor interactions, which map naturally onto superconducting architectures with two-dimensional grid connectivity \cite{fowler2012surface, versluis2017scalable, google2023suppressing, krinner2022realizing}. The surface code has also been adapted to alternative qubit technologies, requiring a dedicated mapping layer to reconcile the code’s 2D connectivity with the physical architecture \cite{pataki2025compiling, veldhorst2017silicon, hetenyi2024tailoring}.

In this work, we take the opposite approach. Rather than starting from a fixed hardware architecture and determining how best to embed a quantum error correction code into it, we synthesize a spin-qubit-based architecture that is designed from the ground up to encode a single logical qubit as efficiently and robustly as possible under realistic architectural constraints. We name this methodology \textbf{Quantum Reverse Mapping}, as it inverts the traditional perspective: rather than adapting the code to a fixed hardware platform, we derive the hardware layout directly from the intrinsic requirements of the error-correcting code, effectively reverse-synthesizing an architecture that optimally supports it.

Looking forward to large-scale quantum processors, the problem shifts from the mapping of a single logical, error-corrected qubit to the compilation of full quantum circuits consisting of many logical qubits. At that scale, optimally compiling logical circuits onto physical layouts is intractable, as the underlying mapping and routing problems are NP-complete \cite{siraichi2018qubit}. As a result, practical compilers rely on heuristic routing strategies and scheduling algorithms that provide good but inherently sub-optimal solutions within reasonable time \cite{li2019tackling, niu2020hardware, escofet2024route}. The effectiveness of such heuristics, however, depends critically on the quality of the logical-qubit encoding at the lowest layer. By establishing an optimized encoding of a single logical qubit, we provide a strong foundation that future heuristic compilation layers can build upon, thereby improving scalability and noise resilience.

On a lower layer, silicon spin qubits \cite{loss1998quantum, wild2012few, vandersypen2019quantum} have emerged as an especially attractive hardware platform for industrially scalable devices; they combine long coherence times \cite{veldhorst2014addressable}, gate fidelities exceeding 99\% \cite{kawakami2016gate, yoneda2018quantum, xue2022quantum}, and compatibility with current semiconductor manufacturing processes, allowing for easier integration with classical control electronics \cite{zwanenburg2013silicon}.

Another interesting property of spin qubits is their native capability for qubit shuttling \cite{mills2019shuttling, jadot2021distant, seidler2022conveyor, PRXQuantum.4.030303}. 
By transporting qubits along 1D channels (often called shuttling tracks or buses) we can dynamically bring qubits into proximity precisely when interactions are required. This mechanism effectively underlies many promising, scalable architectures for semiconductor spin-qubit platforms \cite{buonacorsi2019network, boter2022spiderweb, Patomäki2024, kunne2024spinbus}.

Shuttling operations, as we discuss below, are fundamental to the functionality of future spin-based quantum processors. Due to practical limitations, long-range dot-to-dot coupling across extended quantum devices is not feasible for large-scale architectures. Instead, quantum states must be transferred between distant sites through reliable and coherent state transfer mechanisms.  

Qubit shuttling emerged as one of the most promising mechanisms for silicon spin qubits. It offers a potential pathway to mitigate connectivity constraints. By physically moving an electron or hole within a quantum dot array or tile, shuttling enables flexible routing of quantum information, access to charge sensors and effective long-range connectivity. 
Across some notable works on shuttling architectures for spin qubits, we note the following architectures. Study~\cite{Taylor2005} proposes one of the earliest architectures for quantum computation using semiconductor spins qubits. The architecture proposes to use two adjacent gate-defined double quantum dots, high-frequency microwave gates and capacitive coupling gates to form a node. Alternating gates are used to shuttle the spin, and qubits are connected by a spin shuttle track (“electron pump”). The study proposes a universal set of quantum gates built for singlet-triplet qubits with charge coupling, which facilitates active error suppression, and outlines a scalable architecture with favourable error thresholds for fault-tolerant operation. 

Study~\cite{Hollenberg2006} proposes a 2D donor electron-spin quantum computer architecture enabled by an electron spin transport mechanism. Shuttling of spins enables 2D scaling and is a core architectural feature of the study. Paper~\cite{kunne2024spinbus} introduces SpinBus, an architecture that explicitly uses electron shuttling in Si/SiGe to connect qubits as a scaling strategy. This architecture emphasises that spin shuttling requires low operating frequencies for maintaining qubit coherence. 

Another recent discussion on shuttling architecture is presented in~\cite{Matsumoto2025}. Spin shuttling is used to move qubits into an interaction zone. 
Reference~\cite{Nemeth2024_OmnidirectionalShuttling} addresses a Si/SiGe-specific issue -- valley excitations near valley-splitting minima during shuttling.
This study proposes a 2D shuttling pattern that enables omnidirectional motion for network routing.

Compilation ideas are presented in~\cite{crawford2023compilation}. This study considers a sparse 2D connectivity and proposes a modular 2D spin-qubit architecture built from silicon spin qubits. The paper introduces a silicon “junction” that can couple up to four nanowires into 2D arrangements using spin shuttling and SWAP operations.

Two critical parameters define the performance of any shuttling system: the shuttling duration, which must be significantly shorter than the qubit coherence time, and the amount of additional decoherence introduced during the transfer. In the literature, two primary shuttling modes have been established: bucket-brigade and conveyor-mode. 
In the bucket-brigade mode~\cite{mills2019shuttling, PRXQuantum.4.030303}, a carrier (an electron or a hole) is transferred from dot to dot via tunnelling, which leads to increased sensitivity to charge noise. In contrast, conveyor-mode shuttling~\cite{seidler2022conveyor, Xue2024_QubusConveyor} transports the carrier by smoothly translating the confining potential well, thereby eliminating tunnelling events and associated noise sources~\cite{DeSmet2025}, making it a promising candidate for scalable implementations.

Experimentally, conveyor-mode shuttling is typically implemented by applying phase-shifted, periodic voltages to a sequence of control electrodes~\cite{DeSmet2025, Jeon2025_ArchitecturesPulses}. This creates a moving potential landscape that carries the quantum state across the device. Sinusoidal waveforms are often employed due to their simplicity; they are not necessarily optimal for minimizing decoherence or maximizing shuttling fidelity, and more sophisticated control strategies may offer improved performance~\cite{Sokolov2025,Nagai2025_DigitalControlledConveyor}.

While shuttling provides increased architectural flexibility, different material platforms present distinct advantages and challenges. In Si/SiGe heterostructures, conveyor-mode electron shuttling is generally considered less noisy than bucket-brigade shuttling; however, variations in valley splitting~\cite{Losert2023}, material disorder, potential roughness, wiring density, and heating effects~\cite{langrock2023blueprint} can introduce additional sources of decoherence. In germanium-based structures, hole shuttling benefits from strong spin–orbit coupling but may face different noise mechanisms. In GaAs-based structures, the absence of valley degeneracy allows for potentially smoother shuttling, but the presence of non-zero nuclear spins leads to strong hyperfine interactions, significantly reducing the spin coherence time $T_2^*$ compared to Si/SiGe heterostructures~\cite{Fujita2017, Flentje2017}. In Si/SiGe heterostructures, an isotopically purified Si-28 channel is used to enhance the coherence of the spin, but the small valley splitting energy might be taken into account through different architecture solutions as we discussed above~\cite{Nemeth2024_OmnidirectionalShuttling, DegliEsposti2024}

In this work we synthesize an optimal one‑dimensional shuttling‑bus architecture \cite{kunne2024spinbus, Patomäki2024} for the rotated surface code. Our approach combines a mixed‑integer optimization model for small code distances with a scalable heuristic for larger distances, by exploiting coherent spin shuttling \cite{seidler2022conveyor, PRXQuantum.4.030303}. The resulting architecture minimizes both the total shuttling distance and cycle latency, obtaining improved logical error rates compared to the baseline architecture synthesis method. 

Established layout optimization techniques (including classical standard VLSI placement \cite{pucknell1994basic, sechen1986timberwolf3} and emerging quantum computing approaches \cite{yin2025qecc}) are inapplicable to the unique properties and  constraints of the considered architecture, which leverages 1D global parallel shuttling. Consequently, our synthesis techniques are specifically tailored to the architecture at hand, aligning with the growing paradigm of application-aware co-design of quantum systems, where specific architectures are tailored to algorithmic requirements to maximize efficiency under hardware constraints \cite{li2020towards, tomesh2021quantum, li2021co, lin2022domain}.

The main contributions of this work are as follows:
\begin{itemize}
    \item The development of a mathematical model for optimally synthesizing spin qubits in a shuttle bus architecture, minimizing the error overhead added by the shuttlings needed for the syndrome extraction of the chosen quantum error correction code.
    \item The proposal of an optimal-inspired heuristic placement technique for the scalable synthesis of larger code distances specifically tailored for the surface code.
    \item An analysis of the pseudo-threshold values for each noise source of the considered noise model, consisting of depolarizing noise for quantum gates and dephasing noise for latency-induced errors.
    \item An exploration of the architecture design decisions' impact on the performance of the surface code, considering shuttling velocity, distance between elements, and architecture synthesis techniques.
\end{itemize}

The remainder of the paper is structured as follows. In Section \ref{sec:shuttling_bus_arch} we introduce the considered architecture and review similar work. Section \ref{sec:surface_code_synthesis} explains how the surface code can be mapped into the architecture and explores optimal and heuristic qubit placement. In Section \ref{sec:qec_results} we explore how the proposed layout performs for the surface code, exploring both noise and architecture parameters. The work is concluded in Section \ref{sec:conclusions} where the outcomes of the paper are summarized and future work directions are outlined.

\section{The Shuttling Bus Architecture}
\label{sec:shuttling_bus_arch}

Large‑scale spin‑qubit processors face a severe pin‑count bottleneck: each physical qubit typically demands multiple DC, pulsed, or microwave control lines, so that the required number of classical wires grows with the number of qubits, whereas the available routing cross‑section grows only with the perimeter of the processor \cite{vandersypen2017interfacing}. Architectures such as cross‑bar \cite{li2018crossbar} and bilinear layouts \cite{9720606} alleviate, but do not eliminate, the fan‑out problem.

A promising remedy is to arrange qubits along a 1‑D “conveyor” track and shuttling them to fixed gate regions only when interactions are required. Patomäki \textit{et al.} introduced a pipeline architecture in which qubits advance column‑by‑column through predefined single‑ and two‑qubit gate sites under global timing signals \cite{Patomäki2024}. Künne \textit{et al.} proposed the SpinBus architecture, a modular array of parallel shuttling lanes that provides 2D logical connectivity while requiring only four phase‑shifted sinusoidal voltages per lane \cite{kunne2024spinbus}. Both schemes leverage recent demonstrations of coherent electron‑spin shuttling \cite{seidler2022conveyor, PRXQuantum.4.030303}. 

\begin{figure}
    \centering
    \includegraphics[width=1\linewidth]{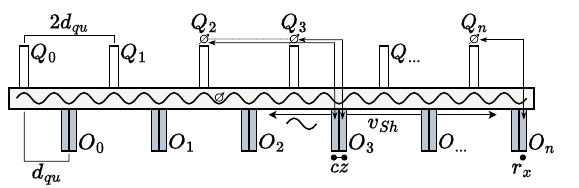}
    \caption{Sketch of the proposed shuttling bus architecture. Spins in quantum dots $\{ Q_0, \ldots, Q_n \}$ are shuttled to manipulation zones $\{ O_0, \ldots, O_n \}$ when they have to interact. The distance between consecutive elements in the 1D shuttling bus is $d_{qu}$, and the shuttling velocity of the spins is set to $v_{sh}$.}
    \label{fig:shuttling_bus_architecture}
\end{figure}

In this work, we adopt a single-lane variant of these proposals, as depicted in Figure \ref{fig:shuttling_bus_architecture}. We call this architecture the \emph{shuttling bus architecture}, which is divided into storage and manipulation zones. Storage zones are quantum dots where qubits reside idly, these storage zones being connected through a shuttling bus mechanism to manipulation zones, which are designated areas for single- and two-qubit gates, as well as readout operations.

A global four‑phase waveform translates the confinement minimum along the track, conveying qubits from storage zones to manipulation zones each time they must interact. Because the waveform is broadcasted, the number of cryogenic control signals is independent of the number of qubits, directly addressing the perimeter‑vs‑area imbalance posed before \cite{seidler2022conveyor, kunne2024spinbus}. 

Qubits can be in storage or manipulation zones. They are moved from one zone to another via the shuttling bus connecting all the zones. This means that, if two qubits are on a track they should move in the same direction (either right or left), and at the same speed. Quantum gates can only be applied to qubits located in manipulation zones. Therefore, each time a gate needs to be executed, the associated qubit must be shuttled from a storage zone to a manipulation zone, where the gate will be applied. Moreover, for the execution of two-qubit gates, the two qubits involved must be in the same manipulation zone.

Previous studies have examined similar systems. In~\cite{cai2023looped} a shuttle based looped-pipeline architecture is used to emulate 3D connectivity within a strictly 2D device for the surface code. More recently,~\cite{yenilen2025performancespinqubitshuttling} also studied the surface code implementation in the SpinBus architecture. In parallel, other works focused on proposing compilation techniques for circuit execution in shuttling-based architectures \cite{escofet2025compilation, crawford2023compilation, paraskevopoulos2024besnake}.

In the following sections, we used the proposed architecture to store a single logical qubit using the surface code, assessing how different layouts (\textit{i.e.}, different distributions of storage and manipulation zones) perform, using both architectural metrics and the logical error rate of the code obtained through simulation.

\section{Shuttling Bus Architecture Synthesis for the Surface Code}
\label{sec:surface_code_synthesis}

\begin{figure}
    \centering
    \includegraphics[width=1\linewidth]{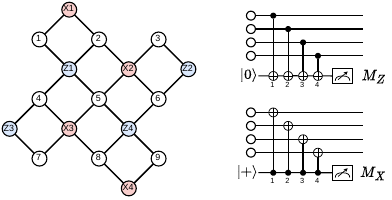}
    \caption{The 2D planer layout of the distance-3 surface code (left). Data qubits are represented with white nodes and $X$ and $Z$ checks are red and blue nodes respectively. Top and bottom circuits (right) are $Z$- and $X$-syndrome extraction circuits, respectively, each consisting on four CNOT gates.}
    \label{fig:surface_code_3}
\end{figure}

The surface code~\cite{bravyi1998quantum, dennis2002topological, fowler2009high, fowler2012surface, litinski2019game, google2023suppressing} is a promising family of 2D topological stabilizer codes defined on a square lattice, featuring constant-weight (4-weight) stabilizer generators and an array of $d^2$ data qubits. The surface code has been widely studied for both spin qubits~\cite{xue2022quantum, cai2023looped}, and other qubit technologies~\cite{google2023suppressing, zhao2022realization, barends2014superconducting, trout2018simulating, kang2023quantum}, due to its 2D planar layout, high threshold values, and the existence of efficient decoding algorithms.

As shown in Figure~\ref{fig:surface_code_3}, the distance‑3 rotated surface code uses 9 data qubits (1 -- 9) and 8 ancilla qubits ($X1$~--~$X4$, and $Z1$~--~$Z4$), each participating in weight-4 stabilizer checks that detect both bit-flip and phase-flip errors. During each QEC cycle, these stabilizers are measured in parallel to extract syndromes, which are fed into the decoder to maintain logical fidelity.

\subsection{Chain Decomposition for Syndrome Extraction}

For $X$ and $Z$ checks to be carried out in parallel while avoiding hook (or horizontal) errors~\cite{dennis2002topological}, each check must be performed in a zig-zag pattern across the data qubits \cite{orourke2024comparepairrotatedvs, tomita2014low, fowler2012surface}. In this work we will use the interaction pattern shown in Figure \ref{fig:hook_errors} \cite{gidney2021stim}.

\begin{figure}
    \centering
    \includegraphics[width=\linewidth]{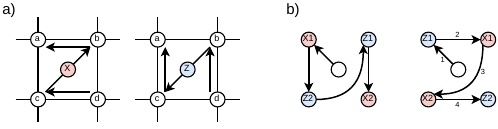}
    \caption{Check interaction order to avoid hook (or horizontal) errors. \textbf{a)} Interaction order from the \textit{check/ancilla qubit point of view} to avoid hook errors. \textbf{b)} Interaction order from the \textit{data qubit point of view} to avoid hook errors. Note that the two figures have equivalent interactions order, depending on whether seen from the check/ancilla or the data qubits point of view. Numbers reference CNOT order as in Figure \ref{fig:surface_code_3}.}
    \label{fig:hook_errors}
\end{figure}

\begin{figure}
    \centering
    \includegraphics[width=0.55 \linewidth]{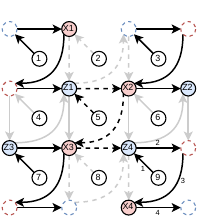}
    \caption{Composite interaction order for the data qubits in the distance-3 surface code lattice. Composite interaction order for the data qubits in the distance-3 surface code lattice. Empty, dashed qubits are shown at the patch boundaries to illustrate how the interaction structure extends when scaling to larger code distances.}
    \label{fig:composite_shuttling}
\end{figure}

When composing the different interaction orders of each data qubit (from Figure \ref{fig:hook_errors}) and grouping them into the full surface code patch (see Figure \ref{fig:surface_code_3}) we obtain a directed graph representing all interactions between data and ancilla qubits for the syndrome extraction of the surface code, as it can be seen in Figure \ref{fig:composite_shuttling}. It is important to notice that the aforementioned graph (called the \emph{composite interaction directed graph}) has no cycles, which will be a critical property for the architecture synthesis method proposed in latter sections.

\begin{figure}
    \centering
    \includegraphics[width=0.7\linewidth]{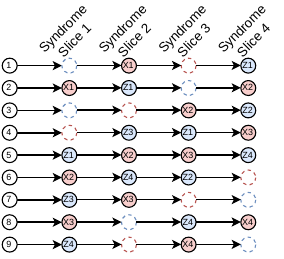}
    \caption{Each data qubit's interactions for each syndrome extraction slice in the distance-3 surface code.}
    \label{fig:time_chains}
\end{figure}

From the interaction pattern depicted in Figure \ref{fig:composite_shuttling} we can take, for each data qubit, the interactions with the different checks in the code and create what we call the \emph{chain interactions}. These data-qubit-wise chain interactions (depicted in Figure \ref{fig:time_chains}) show what each data qubit is doing at each of the four slices of the syndrome extraction circuit. Note that data qubits at the border of the surface code patch have \emph{idling} slices, where they are not interacting with any ancilla qubits from the code.

Due to the acyclic property of the composite interaction directed graph introduced before (see Figure \ref{fig:composite_shuttling}) we can unify all the data qubit chains by merging nodes with the same label (\textit{i.e.}, the ancilla qubits of the code).

By doing this we add to the ancilla qubits the relative order of interactions for data qubits (\textit{e.g.}, data qubits $D1$ and $D2$ interact with $Z1$ after interacting with $X1$, and data qubit $D4$ interacts with $Z1$ after interacting with $Z3$) creating what we call the \emph{chain-preserving directed acyclic graph} or chain-preserving DAG for short. Note that this graph can be constructed thanks to the acyclic property of the composite interaction graph, otherwise we would have cyclic dependencies on the order of the ancilla qubits, and the resulting chain-preserving graph would not be acyclic \cite{thulasiraman19925}.

\begin{figure}
    \centering
    \includegraphics[width=\linewidth]{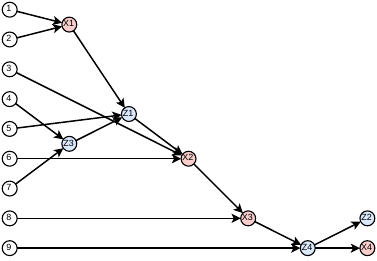}
    \caption{Chain-preserving directed acyclic graph. Each incoming edge to a node represents a dependency on the shuttling bus placement, meaning that if node $A$ is connected to node $B$ ($A \rightarrow B$), $A$ has to be placed left of $B$ in the architecture.}
    \label{fig:dag}
\end{figure}

Figure \ref{fig:dag} shows the chain-preserving DAG for the surface code with distance $d=3$.

\subsection{From the Surface Code to the Shuttling Bus Architecture}

In designing the shuttling bus architecture, we must specify how storage and manipulation zones are arranged along the one-dimensional bus, and determine which data and ancilla qubits are assigned to each zone. While many possible assignments exist, the architecture must be designed so that qubit placement supports the surface code as robustly as possible, minimizing shuttling overhead and maximizing the performance of the logical encoding.

For this, and relying on other works dealing with the same architecture \cite{escofet2025compilation, cai2023looped}, we only explore the placements of qubits that allow for the parallel shuttling of data qubits. With this constraint, which we call the \emph{parallel shuttling constraint}, the shuttling bus will move all data qubits simultaneously into manipulation zones, that will be partly occupied by ancilla qubits.

One could ask the question whether a solution allowing for the parallel shuttling of all qubits exists, since the qubit interactions are given by a 2D code, and we are mapping it into a 1D placement. Fortunately, the chain-preserving DAG constructed before ensures that: each data qubit can be moved only in one direction (\textit{i.e.,} always to the right or left of the shuttling bus) and will meet all the ancilla qubit it should interact with. If a cycle was present on the composite interaction graph, that constraint would not be met, and a solution satisfying the parallel shuttling constraint would not exist.

To demonstrate the importance of this constraint, let us consider the following example. Consider a simple illustrative case with two data qubits $D1$ and $D2$, and two ancilla qubits $A$ and $B$. $D1$ is scheduled to interact first with $A$ and later with $B$. $D2$, on the other hand, is scheduled to interact first with $B$ and later with $A$. If we translate these interaction orders into edges of the chain-preserving directed (acyclic) graph, we obtain a directed path $D1 \rightarrow A \rightarrow B$ from the first sequence and $D2 \rightarrow B \rightarrow A$ from the second one. Because the two paths from the graph point in opposite directions between $A$ and $B$, they close into a cycle $A \rightarrow B \rightarrow A$, thus violating the acyclic property required by the chain-preserving DAG, meaning no parallel-shuttling schedule can satisfy all constraints in this example.

Since the above problem is not present in the surface code, any topological sort of the chain-preserving DAG created before results in a feasible placement of the qubits for the shuttling bus architecture, ensuring that all data qubits are moved in parallel to the right (or left) of the shuttling bus. A topological sort or topological ordering of a directed graph is a linear ordering of its vertices such that for every directed edge $(u,v)$ from vertex $u$ to vertex $v$, $u$ comes before $v$ in the ordering \cite{cormen2022introduction}.

 For this work, we (trivially) assume that data qubits will always be shuttled to the right, though this decision is not important since the results could be mirrored in the other direction.

\begin{figure}
    \centering
    \includegraphics[width=1\linewidth]{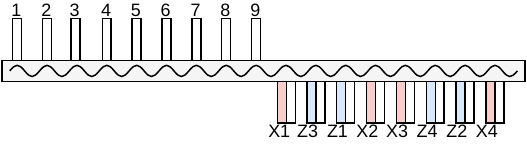}
    \caption{\emph{Naive-topological} placement for the distance-three surface code into the shuttling bus architecture.}
    \label{fig:naive_placement}
    \vspace{-7pt}
\end{figure}

As baseline for our work, we introduce the \emph{Naive-topological} placement of the surface code into the shuttling bus architecture. For this naive layout, we compute a trivial topological sort of the chain-preserving DAG using Kahn's algorithm \cite{kahn1962topological}, and the resulting qubit order is then mapped into the architecture. Figure \ref{fig:naive_placement} shows the resulting layout for the surface code with distance $d=3$.

\subsection{Optimal Architecture Synthesis}
We propose a mathematical model that synthesizes the physical architecture by optimally placing qubits in a one-dimensional shuttling bus. The model aims to obtain the qubit placement (\textit{i.e.,} data and ancilla qubit ordering) that minimize both the total shuttling distance and the time for each error correction cycle. 

Because of the parallel shuttling constraints posed before, all data qubits will be moved in between data-ancilla interactions to the right of the bus. So, at each timestep $t$, only the longest shuttle in that timestep will contribute to the overall latency, effectively hiding the latency of other, shorter, shuttles.

Let $\mathcal{C}$ be the chained interactions for a surface code, with the form $(d, a_1, a_2, a_3, a_4)~\in~\mathcal{C}$, where $d \in \mathcal{D}$ is a data qubit and $a_i \in \mathcal{A}$ is an ancilla qubit of the code. For a given positions $\mathcal{P}$ satisfying the parallel shuttling constraint (\textit{i.e.,} $\mathcal{P}_d < \mathcal{P}_{a_1} < \mathcal{P}_{a_2} < \mathcal{P}_{a_3} < \mathcal{P}_{a_4}$) and allocating both the data and ancilla qubits ($\mathcal{Q} := \mathcal{D} \cup \mathcal{A}$), the total distance $D(\mathcal{P})$ to be shuttled for satisfying all interactions between qubits is given by:

\begin{align}
    D(\mathcal{P}) &= \sum_{c\in \mathcal{C}} \sum_{i} \mathcal{P}_{c_{i+1}} - \mathcal{P}_{c_i}\\
    &= \sum_{(d,a_1,...) \in \mathcal{C}} ( \mathcal{P}_{a_1} - \mathcal{P}_{d} + \\
    & \qquad \qquad \qquad \mathcal{P}_{a_2} - \mathcal{P}_{a_1} + \nonumber\\
    & \qquad \qquad \qquad \mathcal{P}_{a_3} - \mathcal{P}_{a_2} + \nonumber\\
    & \qquad \qquad \qquad \mathcal{P}_{a_4} - \mathcal{P}_{a_3} + \nonumber\\
    & \qquad \qquad \qquad \mathcal{P}_{a_4} - \mathcal{P}_{d} ) \nonumber\\
    &= \sum_{(d,a_1,...) \in \mathcal{C}} 2 (\mathcal{P}_{a_4} - \mathcal{P}_{d})
\end{align}
    
For the same qubit allocation ($\mathcal{P}$), the distance contributing to time $T(\mathcal{P})$ (\textit{i.e.,} the maximum distance to shuttle in each slice becomes the time bottleneck, since other shuttling latencies can be hidden due to the parallel shuttling employed) can be computed as:

\begin{align}
    T(\mathcal{P}) &= \sum_i \max_{c \in \mathcal{C}} \mathcal{P}_{c_{i+1}} - \mathcal{P}_{c_{i}}\\
    &= \max_{(d,a_1,...) \in \mathcal{C}} \mathcal{P}_{a_1} - \mathcal{P}_{d} + \\
    & \quad \max_{(d,a_1,...) \in \mathcal{C}} \mathcal{P}_{a_2} - \mathcal{P}_{a_1} + \nonumber\\
    & \quad \max_{(d,a_1,...) \in \mathcal{C}} \mathcal{P}_{a_3} - \mathcal{P}_{a_2} + \nonumber\\
    & \quad \max_{(d,a_1,...) \in \mathcal{C}} \mathcal{P}_{a_4} - \mathcal{P}_{a_3} + \nonumber\\
    & \quad \max_{(d,a_1,...) \in \mathcal{C}} \mathcal{P}_{d} - \mathcal{P}_{a_4} \nonumber
\end{align}

We formulate the allocation problem as a Mixed-Integer Linear Program (MILP) problem \cite{bertsimas1997introduction, schrijver1998theory}, in which the objective functions and all constraints are linear expressions, the positions of the qubits are encoded in a binary variable $x_{q,p} \in \{0,1\}$, and time constraints for each slice are encoded in integer variables $t_i \in \mathbb{R}_+$. By encoding the positions as binary variables instead of integer variables we reduce the execution time of the solver.

Within this encoding, $x_{q,p} = 1$ when qubit $q \in \mathcal{Q}$ is allocated in position $p$, and $x_{q,p} = 0$ otherwise. Therefore, in order to obtain the position of a qubit $q$ (previously $\mathcal{P}_{q}$), we now need to iterate through all the possible positions:

\begin{equation}
    \mathcal{P}_{q} = \sum_{p \in \mathcal{P}} p x_{q,p}
\end{equation}

The goal of the optimization problem is to concurrently minimize both the \textit{time} and \textit{distance} of the shuttles needed for satisfying all interactions in the syndrome extraction circuit. In the minimization function (see Equation (\ref{eq:minimization})) each term is scaled by a hyperparameter ($T$ and $D$ respectively) to tune the time/distance tradeoff on the optimization problem.

Constraints in Equations (\ref{const:pos_to_single_qubit}) and (\ref{const:qubit_to_single_pos}) ensure that each zone contains exactly one qubit, and that each qubit is allocated into exactly one position. The parallel shuttling constraint, satisfying the chain dependencies between data and ancilla qubits, is satisfied by Equation (\ref{const:chain}). Lastly, constraints in Equations (\ref{const:t1}) to (\ref{const:t5}) are used to define the time for each shuttling slice ($t_1$ -- $t_5$), for which its sum is minimized in the optimization function.

\begin{widetext}
    The MILP problem is formulated as follows:
    \begin{align}
        \texttt{Optimize:}
        & \quad \min_{t, x} \quad T \sum_i t_i + D \sum_{(d,a_1,...) \in \mathcal{C}} 2 \left( \sum_{p \in \mathcal{P}}p  x_{a_4, p} - \sum_{p \in \mathcal{P}}p  x_{d, p} \right) = \\
        & \quad \min_{t, x} \quad T  \sum_i t_i + 2D  \sum_{(d,a_1,...) \in \mathcal{C}} \ \sum_{p \in \mathcal{P}} p \left( x_{a_4, p} - x_{d, p} \right) \label{eq:minimization}\\
        \nonumber\\
        \texttt{Subject to:}
        & \quad \forall_{p \in \mathcal{P}} \quad : \quad \sum_{q \in \mathcal{Q}} x_{q,p} = 1
        \label{const:pos_to_single_qubit}\\
        & \quad \forall_{q \in \mathcal{Q}} \quad : \quad \sum_{p \in \mathcal{P}} x_{q,p} = 1
        \label{const:qubit_to_single_pos}\\
        & \quad \forall_{(d,a_1,...) \in \mathcal{C}} \quad : \quad \sum_{p_\in \mathcal{P}} p  x_{d,p} < \sum_{p_\in \mathcal{P}} p  x_{a_1,p} < \sum_{p_\in \mathcal{P}} p  x_{a_2,p} < \sum_{p_\in \mathcal{P}} p  x_{a_3,p} < \sum_{p_\in \mathcal{P}} p  x_{a_4,p}
        \label{const:chain}\\
        & \quad \forall_{(d,a_1,...) \in \mathcal{C}} \quad : \quad t_1 \geq \sum_{p \in \mathcal{P}} p (x_{a_1,p} - x_{d,p})
        \label{const:t1}\\
        & \quad \forall_{(d,a_1,...) \in \mathcal{C}} \quad : \quad t_2 \geq \sum_{p \in \mathcal{P}} p (x_{a_2,p} - x_{a_1,p})
        \label{const:t2}\\
        & \quad \forall_{(d,a_1,...) \in \mathcal{C}} \quad : \quad t_3 \geq \sum_{p \in \mathcal{P}} p (x_{a_3,p} - x_{a_2,p})
        \label{const:t3}\\
        & \quad \forall_{(d,a_1,...) \in \mathcal{C}} \quad : \quad t_4 \geq \sum_{p \in \mathcal{P}} p (x_{a_4,p} - x_{a_3,p})
        \label{const:t4}\\
        & \quad \forall_{(d,a_1,...) \in \mathcal{C}} \quad : \quad t_5 \geq -1 \sum_{p \in \mathcal{P}} p (x_{d,p} - x_{a_4,p}) = \sum_{p \in \mathcal{P}} p (x_{a_4,p} - x_{d,p})
        \label{const:t5}
    \end{align}
\end{widetext}

Note that, in $t_5$, each data qubit $d$ in $(d, a_1, a_2, a_3, a_4) \in \mathcal{C}$ is shuttled from $\mathcal{P}_{a_4}$ to $\mathcal{P}_{d}$ (\textit{i.e.}, from right to left in the shuttling bus). Since $\mathcal{P}_{a_4} > \mathcal{P}_{d}$ by Equation (\ref{const:chain}), to obtain the distance units for such shuttle we need to multiply by $-1$ the distance from $\mathcal{P}_{a_4}$ to $\mathcal{P}_{d}$, thus obtaining the right expression in Equation (\ref{const:t5}).

To solve the MILP problem and obtain the \emph{Optimal} architecture placement, we use the Gurobi solver \cite{gurobi}, which for linear programming problems uses several concurrent optimization algorithms such as the simplex \cite{dantzig1948programming, dantzig1949programming} or dual simplex algorithm \cite{lemke1954dual}, the branch and bound method \cite{land2009automatic}, and interior-point (or barrier) methods \cite{dikin1967iterative, karmarkar1984new}.

\begin{figure}
    \centering
    \includegraphics[width=1\linewidth]{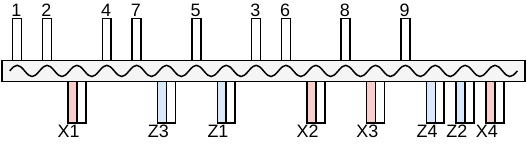}
    \caption{\emph{Optimal} placement for the distance-3 surface code into the shuttling bus architecture.}
    \label{fig:optimal_placement}
\end{figure}

Because silicon spin qubits suffer from low dephasing times \cite{losert2024strategies}, we tune the proposed model so that it prioritizes time over distance in the optimization function by setting $T \gg D$. The resulting optimal shuttling bus architecture for the distance-three surface code is depicted in Figure \ref{fig:optimal_placement}.

\subsubsection*{Quantum Reverse Mapping for Other QEC Codes}
The MILP formulation underlying the Quantum Reverse Mapping framework can be generalized to multiple families of QEC codes, since it does not rely on any code-specific structure beyond the scheduling constraints of the syndrome-extraction circuit. The only requirement for applying the MILP model to an arbitrary quantum error-correcting code is that the ancilla-data interactions used during stabilizer measurement can be represented as a directed acyclic graph (DAG). This ensures that all qubits participating in a stabilizer measurement can be shuttled in the same direction in parallel without introducing cyclic dependencies in the interaction chain (see Figure \ref{fig:dag}).

Once the DAG is available, adapting the optimization model to a new code involves two straightforward modifications.
The ordering constraints that enforce $\mathcal{P}_{d} < \mathcal{P}_{a_1} < \mathcal{P}_{a_2} < \ldots$ (Equation (\ref{const:chain}) for the surface code) must be adjusted according to the ancilla–data interaction chains of the new QEC code. And, the number of time-slice variables ($t_i$) must be adjusted to the maximum stabilizer weight. For the rotated surface code, all stabilizers have weight 4 (with lower-weight boundary checks), so each measurement circuit decomposes into four interaction slices plus a final return step, resulting in five shuttling intervals. Codes with higher-weight stabilizers (e.g., weight-6 color-code checks) naturally require more time variables.

No other part of the MILP formulation requires modification. All objective terms and constraints related to minimizing total shuttling distance and cycle duration remain identical.

To illustrate the generality of the framework, Appendix \ref{sec:color_codes} applies it to the family of 2D color codes \cite{bombin2006topological, landahl2011fault}, a prominent alternative to the surface code due to their lower qubit overhead and ability to perform transversal Clifford gates. These codes feature weight-6 stabilizers and require adapted scheduling and chain extraction, but remain fully compatible with the DAG-based modeling approach.

\subsection{Heuristic Architecture Synthesis}

As the code distance increases, the size of the mathematical optimization model proposed in the previous section grows substantially, both in terms of variables and constraints, making it impractical to solve exactly within a reasonable time frame for larger systems.

To quantify this complexity, consider that for a surface code with distance $d$, the system comprises $N = d^2 + (d^2-1) = O(d^2)$ qubits. The resulting MILP formulation requires defining the problem state using $5$ integer variables ($t_i$) and $N^2$ binary variables $x_{q,p}$ (encoding the mapping of qubits to physical sites). Furthermore, the model necessitates $6N$ linear constraints to ensure correctness. While the number of constraints grows quadratically with $d$, the number of binary variables scales with the fourth power of $d$ ($N \times N = O(d^4)$). Crucially, because the worst-case time complexity of general MILP solvers is exponential with respect to the number of variables (scaling as $O(2^{N^2}) = O(2^{d^4})$  \cite{papadimitriou1981complexity}), this explosion in the search space creates a severe bottleneck for exact solvers as the code distance increases.

Therefore, in order to handle higher code distances, we need a scalable, computationally efficient method. A straightforward option is to generate the topological ordering of qubits using the chain-preserving DAG, the \emph{Naive-topological} placement introduced in earlier sections.

Taking inspiration from the solution obtained using the mathematical model, what we call the \emph{Optimal} placement, we devise a heuristic strategy to place the qubits with the goal of obtaining a similar performance to the \emph{Optimal} placement, while maintaining the computational complexity of the \emph{Naive-topological} one.

\begin{figure}
    \centering
    \includegraphics[width=1\linewidth]{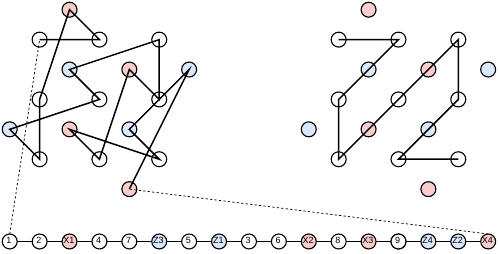}
    \caption{Data qubits order (right) following the \textit{zig-zag} order. The left and bottom diagrams depict the order of all qubits from the surface code, adding ancilla qubits as soon as they are available while satisfying the topological order given by the chain-preserving DAG.}
    \label{fig:order_to_placement}
\end{figure}

To achieve this, we propose the \emph{Zig-Zag} placement. This architecture synthesis strategy starts by placing the top left data qubit of the surface code (\textit{i.e.}, data qubit 1 in Figure \ref{fig:surface_code_3}), and continue by placing the other data qubits in a zig-zag pattern, as depicted in Figure \ref{fig:order_to_placement} (right). Ancilla qubits are placed as soon as they are available satisfying the topological order given by the chain-preserving DAG. The pseudo-code for the proposed synthesis method is given in Algorithm \ref{alg:zig_zag}.

For the distance-3 surface code, the \emph{Zig-Zag} placement strategy exactly matches the \emph{Optimal} one, resulting in the architecture depicted in Figure \ref{fig:optimal_placement}.

\begin{algorithm}[H]
\caption{Zig-Zag Placement Synthesis}
\begin{algorithmic}[1]
\Require Code distance $d$
\Ensure Qubit Placement $\mathcal{P}$

\Statex \textbf{1. Initialization}
\State $\mathcal{D}, \mathcal{A} \leftarrow \texttt{GenerateSurfaceCode}(d)$
\State $G \leftarrow \texttt{ChainPreservizngDAG}(\mathcal{D}, \mathcal{A})$ 

\Statex \textbf{2. Construct Zig-Zag Order}
\State $L \gets$ Data qubits $\mathcal{D}$ ordered in the zig-zag pattern

\Statex \textbf{3. Placement Loop}
\State $\mathcal{P} \gets \emptyset$
\For{each $q_d$ in $L$}
    \State Append $q_d$ to $\mathcal{P}$
    \State Remove $q_d$ from $G$
    
    \State $\mathcal{A}' \gets \{a \in \mathcal{A} \mid \texttt{InDeg}(G, a) =0 \}$
    \While{$\mathcal{A}' \neq \emptyset$}
        \State Sort $\mathcal{A}'$ by priority (connectivity weight)
        \For{$q_a$ in $\mathcal{A}'$}
            \State Append $q_a$ to $\mathcal{P}$
            \State Remove $q_a$ from $G$
        \EndFor
    \EndWhile
\EndFor

\State \Return $\mathcal{P}$
\end{algorithmic}
\label{alg:zig_zag}
\end{algorithm}

The computational complexity of the proposed Zig-Zag placement strategy is dominated by the number of qubits $N$, scaling linearly as $O(N)$ or equivalently $O(d^2)$ with respect to the code distance $d$. The construction of the $L$ order requires visiting each data qubit exactly once, an $O(N)$ operation. By maintaining a map of data-ancilla dependencies (\textit{i.e.}, which ancilla qubits ($\leq 4$) depend on each data qubit) checking and updating the set of available ancillas ($\mathcal{A}'$) inside the main loop can be performed in constant time, independent of the total system size. Consequently, as the placement loop iterates through the data qubits exactly once and performs only local, constant-time updates, the overall computational cost of the synthesis algorithm is $O(N)=O(d^2)$.

\subsection{Comparing the Optimal and Heuristic Architecture Synthesis}

The three proposed architecture synthesis methods, \emph{Naive-topological}, \emph{Zig-Zag}, and \emph{Optimal} placement, only compute the relative positions of the qubits based on the interactions, but the methods are agnostic of the exact timing of each interaction. Since some data qubits placed at the edge of the surface code patch are idling at some syndrome extraction slices, when putting together the chain interactions in the synthesized architecture we may need to add extra spaces for the data qubits to be stored when not interacting. Extra quantum dots to hold the spins in idling slices are added greedily upon necessity.

The \emph{Optimal} and \emph{Zig-Zag} placements alternates data and ancilla qubits in the shuttle bus, creating many feasible positions for each idling data qubit to be stored in. However, the \emph{Naive-topological} method places all data qubits at the beginning (left) of the shuttle bus array, thus not having free intermediate positions for the idling data qubits, thus necessitating more extra quantum dots.

\begin{figure}
    \centering
    \begin{subfigure}[b]{\linewidth}
        \centering
        \includegraphics[width=\textwidth]{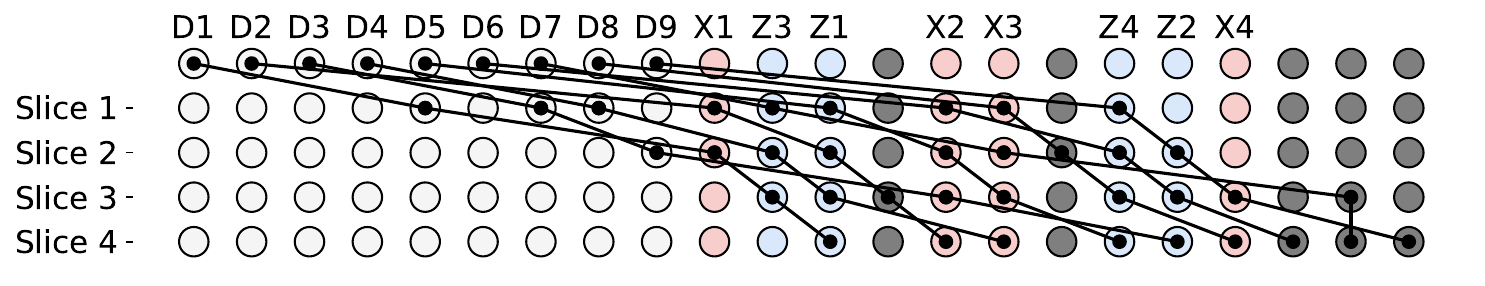}
        \caption{Shuttlings and interactions for the \emph{Naive-topological} placement architecture.}
        \label{fig:naive_shuttlings}
    \end{subfigure}
    \hfill
    \begin{subfigure}[b]{\linewidth}
        \centering
        \includegraphics[width=\textwidth]{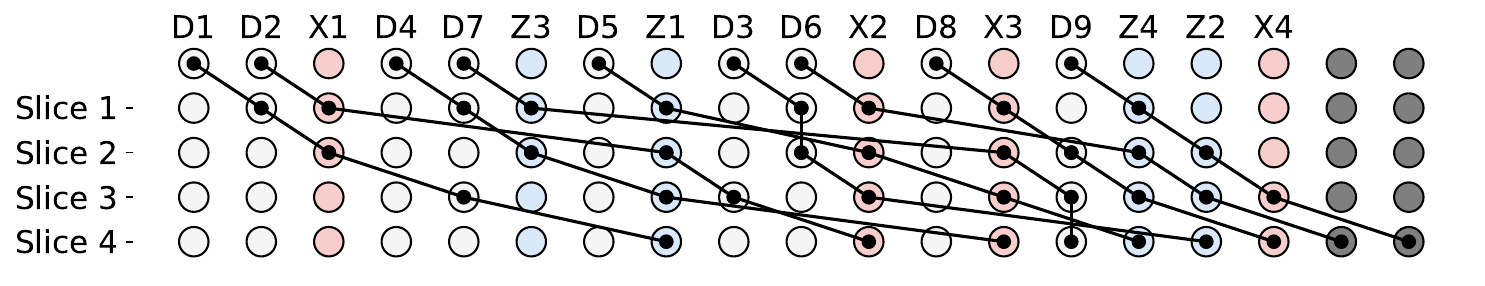}
        \caption{Shuttlings and interactions for the \emph{Optimal} (and \emph{Zig-Zag}) placement architecture.}
        \label{fig:zig_zag_shuttlings}
    \end{subfigure}
    \caption{\emph{Naive-topological} and \emph{Optimal} (and \emph{Zig-Zag}) architectures for the surface code with distance $d=3$. Note that five extra quantum dots are needed for the \emph{Naive-topological} placement, which grows with code distance. On the other hand, for the \emph{Optimal} placement (and also for the \emph{Zig-Zag} one) only two extra quantum dots are needed, which is constant across all code sizes.}
    \label{fig:hook_errors}
\end{figure}

Figures \ref{fig:naive_shuttlings} and \ref{fig:zig_zag_shuttlings} show the shuttles needed of each data qubit across the syndrome extraction circuit for the surface code with distance $d=3$ using the \emph{Naive-topological} and \emph{Optimal} placements respectively. After the last interaction, each data qubit is returned to its original position (shuttle not shown in the figure). Since (for the distance-three surface code) the placements obtained with the \emph{Optimal} and \emph{Zig-Zag} strategies are the same, the shuttlings and interactions between both placements are also the same (see Figure \ref{fig:zig_zag_shuttlings}).

The proposed heuristic \emph{Zig-Zag} placement coincides with the \emph{Optimal} one for low-distance codes. Because of the high computational cost of the mathematical optimization model, we cannot obtain the \emph{Optimal} placement for code distances equal or higher than $d=9$. Therefore, we cannot asses the optimality of the heuristic method for higher code distances. Nonetheless, we can characterize the scaling trends, in terms of shuttling distance, shuttling time, and shuttle bus size (accounting for the extra quantum dots) for the different proposed placement techniques.

To understand the efficacy of the Zig-Zag heuristic, it is useful to view the problem through the lens of local neighborhood preservation. In the Surface Code grid, data qubits that share a common ancilla are always located at distance $<2$ (immediate spatial neighbors). The heuristic aims to preserve this distance-1 connectivity in the linear layout. For example, in a distance-3 code, starting with data qubit 1 at position $(0,0)$, the set of immediate neighbors is $\{(1,0), (0,1), (1,1)\}$ (i.e. data qubits 2, 4, and 5). By prioritizing the placement of $(1,0)$ next to $(0,0)$, the heuristic ensures that the X-type ancilla shared between them (qubit X1 in the example) can be executed with minimal shuttling overhead. Subsequently, the heuristic targets the next available distance-1 neighbor, $(0,1)$, prioritizing it over $(1,1)$ to minimize the number of ancilla qubits added to the 'pending' set (those not yet placed but whose data qubits are already in the placement) thereby preventing the high shuttling distances associated with maintaining these long-range dependencies. This recursive greedy selection of immediate grid neighbors generates the zig-zag pattern. While this is not a formal proof of optimality for all code distances, our MILP results confirm that this geometric strategy yields mathematically optimal placement for code distances $d=3, 5, 7$.

\begin{figure*}
    \centering
    \includegraphics[width=1\linewidth]{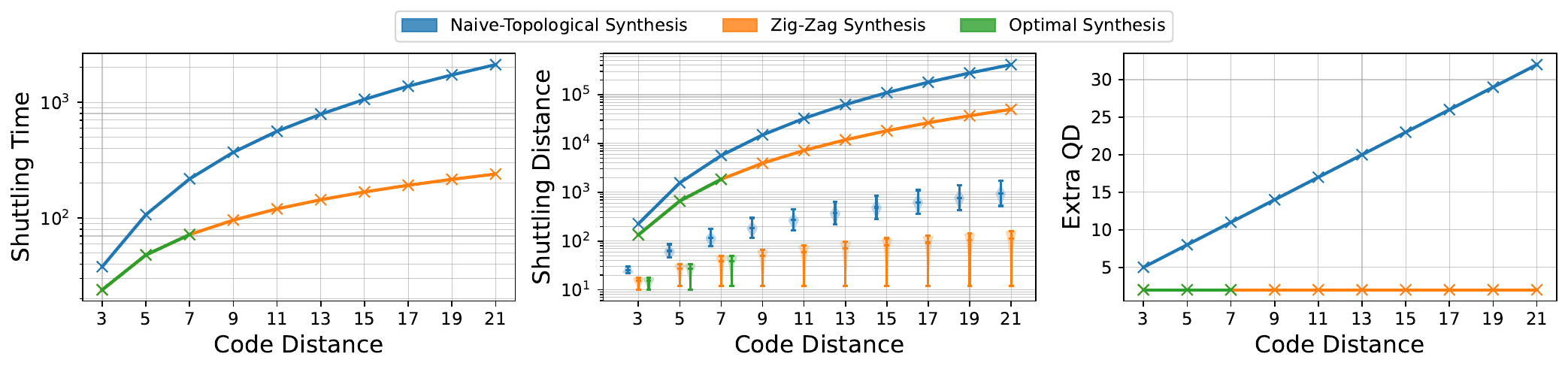}
    \caption{Shuttling time (left), shuttling distance (middle), and added number of extra quantum dots (right) when increasing code distance. Shuttling time (left) can be decomposed into the time of the four different shuttles (one per each slice) and the time of the return shuttle which together form an error correction cycle. The total shuttling time scales linearly for the \emph{Optimal} and \emph{Zig-Zag} architectures, and quadratically for the \emph{Naive-topological} one. Shuttling distance (middle) scales polynomially for all the placement methods. Violin plots depict the variability of the shuttling distance across all data qubits. The number of extra quantum dots (right) that need to be added is constant (2) for the \emph{Optimal} and \emph{Zig-Zag} architectures, and linear for the \emph{Naive-topological} one.}
    \label{fig:scaling_trends}
\end{figure*}

Figure \ref{fig:scaling_trends} depicts the scaling trends for the different proposed architectures, considering code sizes from $d=3$ to $d=21$. The results confirm the \textit{zig-zag} synthesis methodology as a promising candidate, matching the performance of the \textit{optimal} one (for code distance 3, 5, and 7) while maintaining the computational cost of the \emph{Naive-topological} one.

\section{Surface Code Simulation in a Shuttling Bus Architecture}
\label{sec:qec_results}

Beyond evaluating architectural metrics related to scalability, it is essential to assess how well the proposed architecture performs in its intended role: reliably storing a logical qubit using the surface code.

In this section we simulate the surface code memory circuit under realistic noise conditions on the shuttling bus architecture using the \emph{Zig-Zag} placement. For the simulation we use the Stim library \cite{gidney2021stim, 10.1145/3505637}, which has been proven to be an efficient and scalable library for stabilizer circuits simulation.

\subsection{Noise Model}
We assume three sources of error contributing to the noise added to the quantum states, dividing the error sources into idling, gate, and shuttling-induced.

Under realistic conditions, dephasing is expected to be the dominant form of decoherence for silicon spin qubits \cite{losert2024strategies}. Spins that are idling in a quantum dot will be decohered according to $T_2^*$, modeled as a phase damping channel \cite{nielsen2010quantum}. Therefore, a quantum state $\rho$ that stays for $t$ seconds in a quantum dot with decoherence time ${T_2^*}_{QD}$, will be dochered according to:
\begin{align}
    \mathcal{N}(\rho) \rightarrow E_0 \rho E_0^\dagger + E_1 \rho E_1^\dagger
\end{align}

With $E_0 = \begin{bmatrix} 1 & 0 \\ 0 & \sqrt{1-\lambda} \end{bmatrix}$, $E_1 = \begin{bmatrix} 0 & 0 \\ 0 & \sqrt{\lambda} \end{bmatrix}$, and $\sqrt{1-\lambda} = e^{-t/2 {T_2^*}_{QD} }$, which can be modeled with a $Z$-Pauli $\mathcal{{E}}_p (\rho)$ channel \cite{nielsen2010quantum} by setting:
\begin{equation}
    p = \frac{1-e^{-t/2{T_2^*}_{QD}}}{2}
    \label{eq:p_z_channel}
\end{equation}

Allowing us to add it into the Stim simulation, which only accepts Pauli channels as noise sources.

In our simulations, we assume gate times of 100 ns for single-qubit operations and 50 ns for two-qubit operations \cite{langrock2023blueprint, losert2024strategies, ginzel2024scalable, yenilen2025performancespinqubitshuttling}. While in practice single- and two-qubit gates often exhibit different fidelities, we base our model on state-of-the-art results where their performance is comparable, and approximate both as depolarizing channels with parameter $p$. For qubit readout, we assume a duration of 500 ns, reflecting an optimistic yet experimentally supported estimate \cite{ibberson2021large, higginbotham2014coherent, barthel2010fast, stano2022review}, and neglect state-preparation-and-measurement (SPAM) errors. We emphasize that this corresponds to an optimistic assumption for both gate and readout performance, and in the following sections we separately explore the effect of depolarizing noise and coherence times.

It is still unclear how errors due to shuttling affect qubits, though several works assessed it from both theoretical and experimental perspectives \cite{langrock2023blueprint, struck2024spin, losert2024strategies}.

\begin{figure}
    \centering
    \includegraphics[width=1\linewidth]{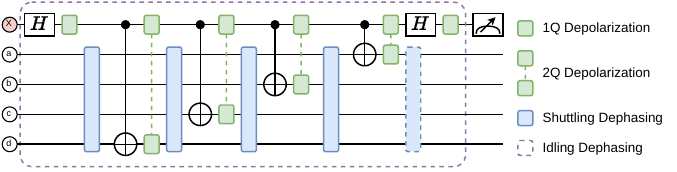}
    \caption{Noisy circuit for the $X$-Checks of the surface code.}
    \label{fig:nosy_x_circuit}
\end{figure}

\begin{figure*}
    \centering
    \includegraphics[width=1\linewidth]{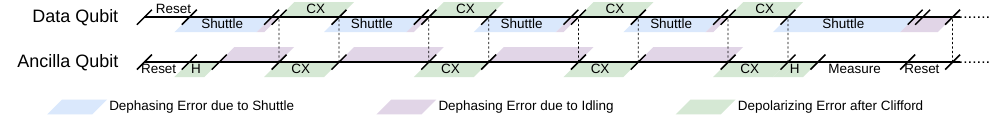}
    \caption{Syndrome extraction cycle breakdown for data and ancilla qubits. Including synchronization steps (vertical dashed lines) and noise applied to each one of the steps.}
    \label{fig:cycle_breakdown}
\end{figure*}

We take the experimental noise characterization obtained in \cite{struck2024spin} and model the shuttling noise as a phase damping channel, similar than for the $T_2^*$ time mentioned before. However, as described in \cite{struck2024spin}, the effective dephasing time for the shuttled states depends on the base dephasing time of the shuttling bus (${T_2^*}_{bus}$), the distance of the shuttle $d_{sh}$, and the quantum dot correlation length $l_c$, resulting in an effective dephasing time ${T_2^*}_{sh}$ for the shuttling given by:
\begin{equation}
    {T_2^*}_{sh} = {T_2^*}_{bus} \sqrt{\frac{d_{sh}+l_c}{l_c}}
    \label{eq:t2_shuttling}
\end{equation} 
which will also be modeled as a phase damping ($Z$-Pauli) noise channel using Equation (\ref{eq:p_z_channel}).

Figure \ref{fig:nosy_x_circuit} depicts the error locations for the $X$ parity checks.

\subsection{Noise Threshold Calculation}
We now want to evaluate how much noise the architecture can tolerate from each one of the three noise sources introduced before. Figure \ref{fig:cycle_breakdown} depicts the steps needed for the syndrome extraction for both the data and ancilla qubits. It also shows which quantum gates need to be synchronized (vertical dashed lines), and what type of noise is applied at each step.

Table \ref{tab:sim_params} summarizes the simulation parameters. Note that $p$ (depolarizing noise), ${T_2^*}_{QD}$, and ${T_2^*}_{bus}$ are set to 0, $+\infty$, and $+\infty$, respectively, since we begin the simulations by analyzing independently each noise source.

\begin{table}
    \centering
    \begin{tabular}{c|c}
         \textbf{Parameter} & \textbf{Value} \\ \hline\hline
         ~Correlation Length ($l_c$) ~&~ 13 nm \cite{langrock2023blueprint} ~\\ \hline
         ~Distance Between Qubits ($d_{qu}$) ~&~ 100 nm~\\ \hline
         ~Shuttling Velocity ($v_{sh}$) ~&~ 2.8 m/s \cite{struck2024spin} ~\\ \hline\hline
         ~Single-Qubit Gate Time ~&~ 100 ns ~\\ \hline
         ~Two-Qubit Gate Time ~&~ 50 ns ~\\ \hline
         ~Measurement Time ~&~ 500 ns ~\\ \hline\hline
         ~Gate Depolarization ($p$) ~&~ 0, 0.001, 0.005 ~\\ \hline
         ~$T_2^*$ Quantum Dot (${T_2^*}_{QD}$) ~&~ $100\mu s$, $1000\mu s$, $+\infty$ ~\\ \hline
         ~$T_2^*$ Shuttling Bus (${T_2^*}_{bus}$) ~&~ [$1\mu s$, $100\mu s$], $+\infty$ ~\\ \hline\hline
         Code Distance ~&~ 3 -- 9 ~\\ \hline
         Synthesis Method ~&~ \emph{Zig-Zag} Placement ~\\ \hline
    \end{tabular}
    \caption{Simulation parameters.}
    \label{tab:sim_params}
\end{table}

To do so, we set two of the three noise sources to lossless values and sweep the third one for different code sizes, setting the number of correction rounds to three times the code distance. From the obtained logical error rate per round, we compute the pseudo-threshold as the mean crossing point across all code distances, also taking into account the standard deviation in the crossings.

\begin{figure}
    \centering
    \includegraphics[width=\linewidth]{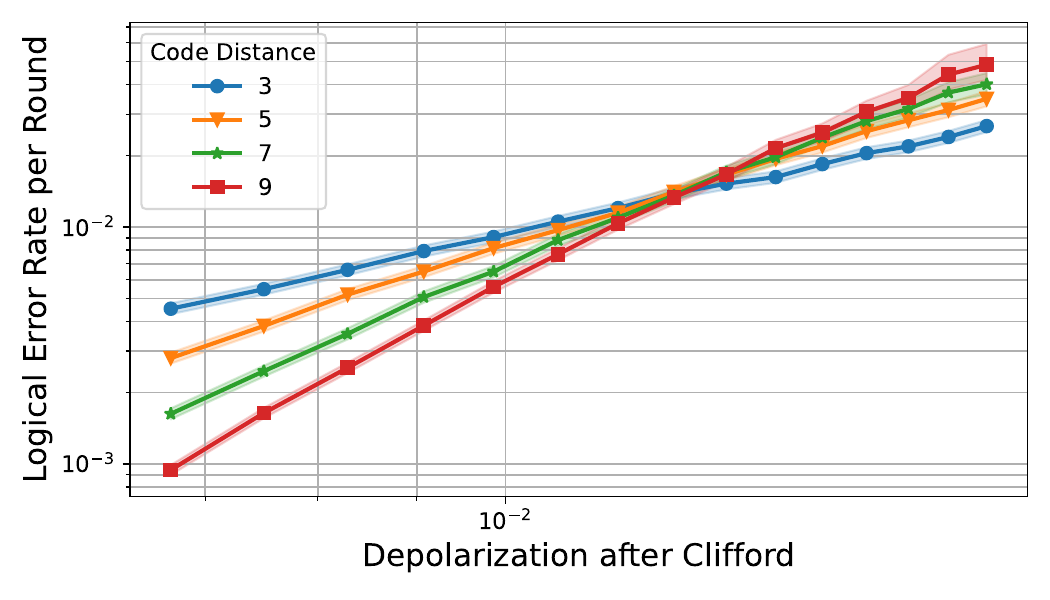}
    \caption{Gate depolarizing noise threshold calculations. Threshold value: $0.012293 \pm 6.35\times10^{-4}$}.
    \label{fig:depol_threshold}
\end{figure}

\begin{figure}
    \centering
    \includegraphics[width=\linewidth]{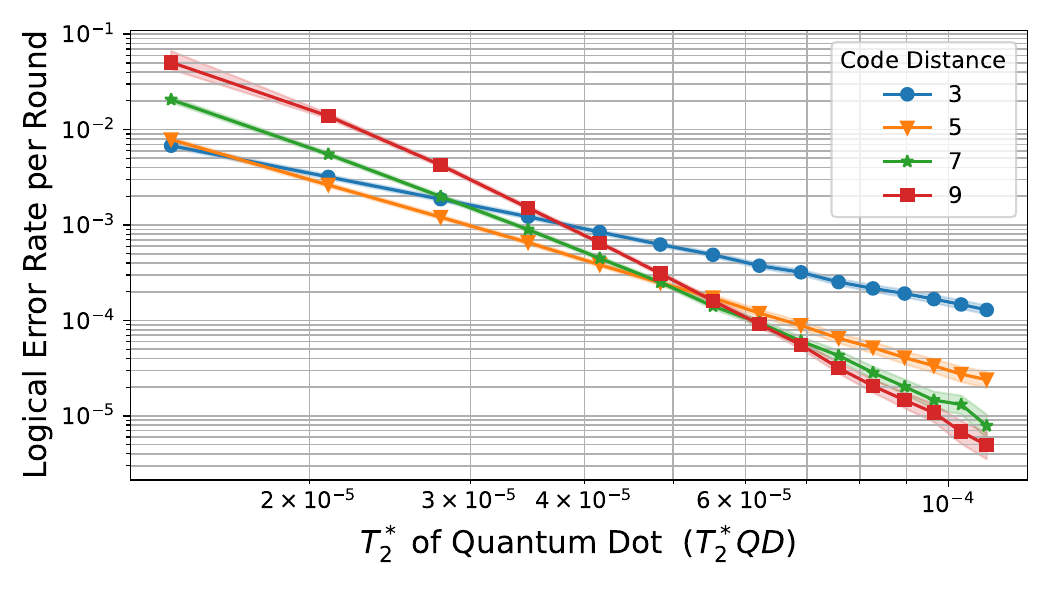}
    \caption{$T_2^*$ of the Quantum Dot noise threshold calculations. Threshold value: $4.94469\times10^{-5} \pm 7.74\times10^{-6}$}.
    \label{fig:t2qd_threshold}
\end{figure}

\begin{figure}
    \centering
    \includegraphics[width=\linewidth]{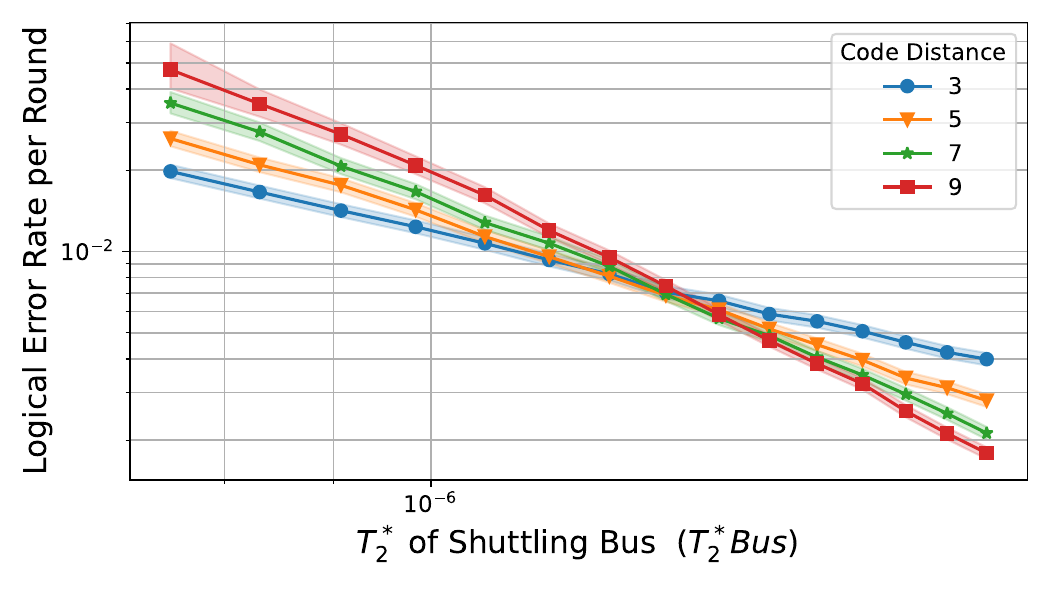}
    \caption{$T_2^*$ of the Shuttling Bus noise threshold calculations. Threshold value: $1.375675\times10^{-6} \pm 7.2\times10^{-8}$}.
    \label{fig:t2bus_threshold}
\end{figure}

Figures \ref{fig:depol_threshold}, \ref{fig:t2qd_threshold}, and \ref{fig:t2bus_threshold} show the logical error rate per round for code distances 3 to 9 when exploring the three different noise sources considered. These figures also report the mean and deviation of the pseudo-threshold.

We can see how, when exploring the different noise sources, the threshold deviation (\textit{i.e.}, whether or not all logical rates distances cross at the same point) varies significantly, specially for the idling dephasing time ${T_2^*}_{QD}$.

When assessing each noise source independently, we obtain upper bounds on how much can the architecture tolerate from each noise source. However, all noises are present in a realistic architecture, and therefore we must asses how the different noises combine together.

\begin{figure*}
    \centering
    \includegraphics[width=1\linewidth]{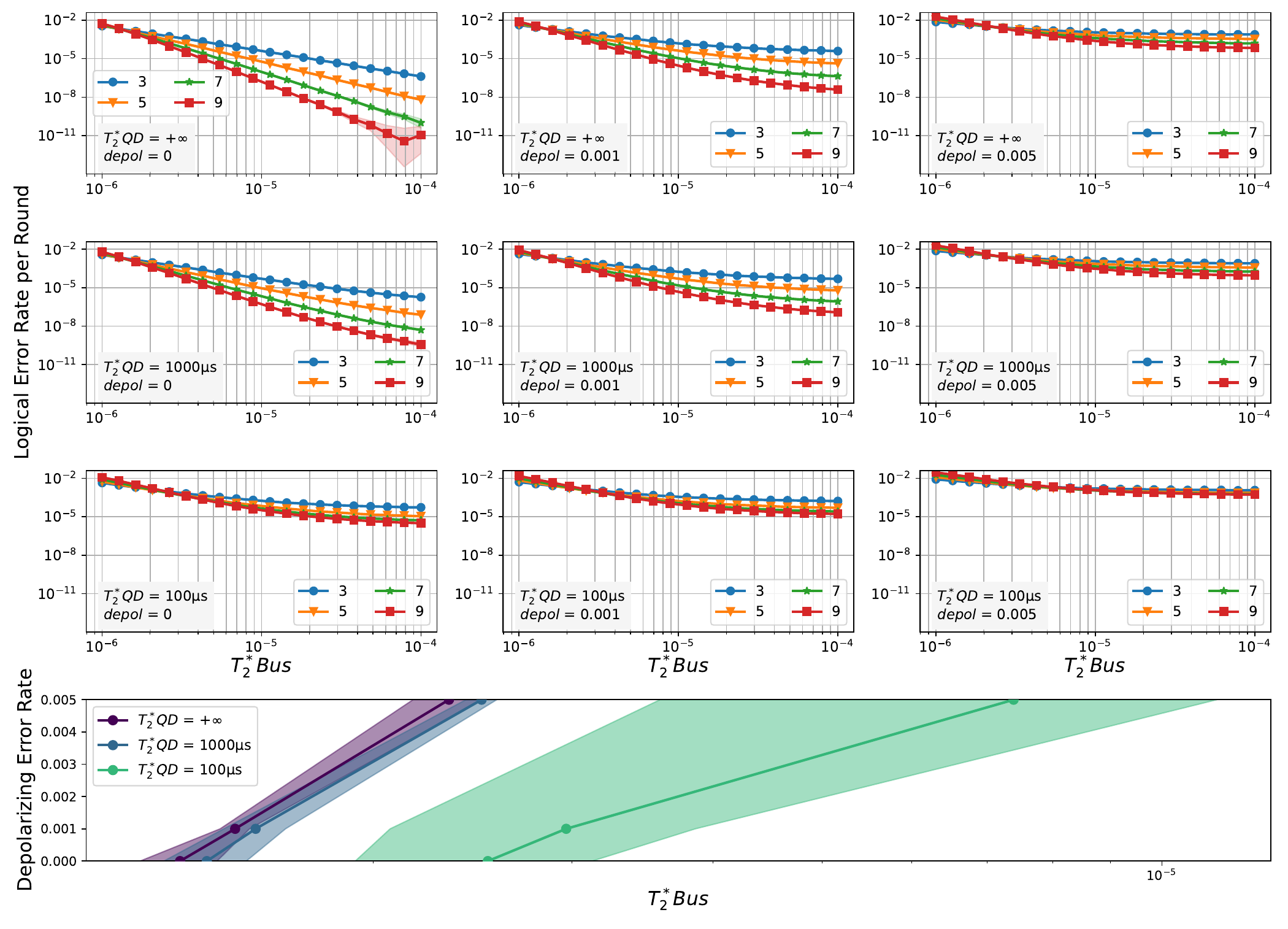}
    \caption{$T_2^*$ of the Shuttling Bus noise threshold calculations under several configurations of depolarizing noise after each gate, and $T_2^*$ of the Quantum Dot.}
    \label{fig:pairwise_thresholds}
\end{figure*}

Since in this work we mainly focus on the shuttling process and its impact on the overall logical error rate, we study how much ${T_2^*}_{bus}$ can the architecture tolerate for some fixed depolarization ($p$) and dephasing times (${T_2^*}_{QD}$). The results are depicted in Figure \ref{fig:pairwise_thresholds}, including, in the bottom row, the threshold trends for the different noise combinations considered.

In the next sections, we explore the impact of the architecture design decisions in the performance of the error correction code by fixing the noise parameters to values above the threshold for the architecture parameters considered until now.

\subsection{Designing Shuttling Bus Architectures}

The previously explored parameters (\textit{i.e.}, the physical noise sources) are intrinsic to the underlying technology and not adjustable by architecture designers. In this section, we shift focus to architectural decisions and analyze their impact on the logical error rate. To isolate these effects, we fix the noise parameters to $p=0.001$ (corresponding to a gate fidelity of 99.925\%), ${T_2^*}_{QD} = 100 \mu s$, and ${T_2^*}_{bus} = 10 \mu s$, and explore how the distance between the elements in the shuttling bus $d_{qu}$ (previously set to $d_{qu} = 100$nm) and the shuttling velocity $v_{sh}$ (previously $v_{sh}=2.8$m/s) affect performance. The value for ${T_2^*}_{QD}$ was chosen under the assumption that quantum dots $\{ Q_0,\ldots,Q_n \}$ (see Fig.~\ref{fig:shuttling_bus_architecture}) do not have micromagnets nearby, which leads to higher ${T_2^*}_{QD}$ due to low influence of charge noise on dephasing~\cite{Chan2021,stano2022review}. We assume that the micromagnets are present only near the dots $\{ O_0,\ldots, O_n \}$, so by shuttling towards any of these dots, in its vicinity, the qubit will experience the magnetic field gradient together with the noise induced by shuttling itself, which as we assume leads to smaller values of ${T_2^*}_{bus}$. However, we do vary ${T_2^*}_{QD}$ and ${T_2^*}_{bus}$ in a wide range of values and study their effect on the code performance.

Because of the effect in ${T_2^*}_{sh}$ of motional narrowing when shuttling across longer distances \cite{struck2024spin} (see Equation (\ref{eq:t2_shuttling})), both parameters $d_{qu}$ and $v_{sh}$ do not have a linear relationship, and each combination of parameters must be tested independently.

\begin{figure*}
    \centering
    \includegraphics[width=1\linewidth]{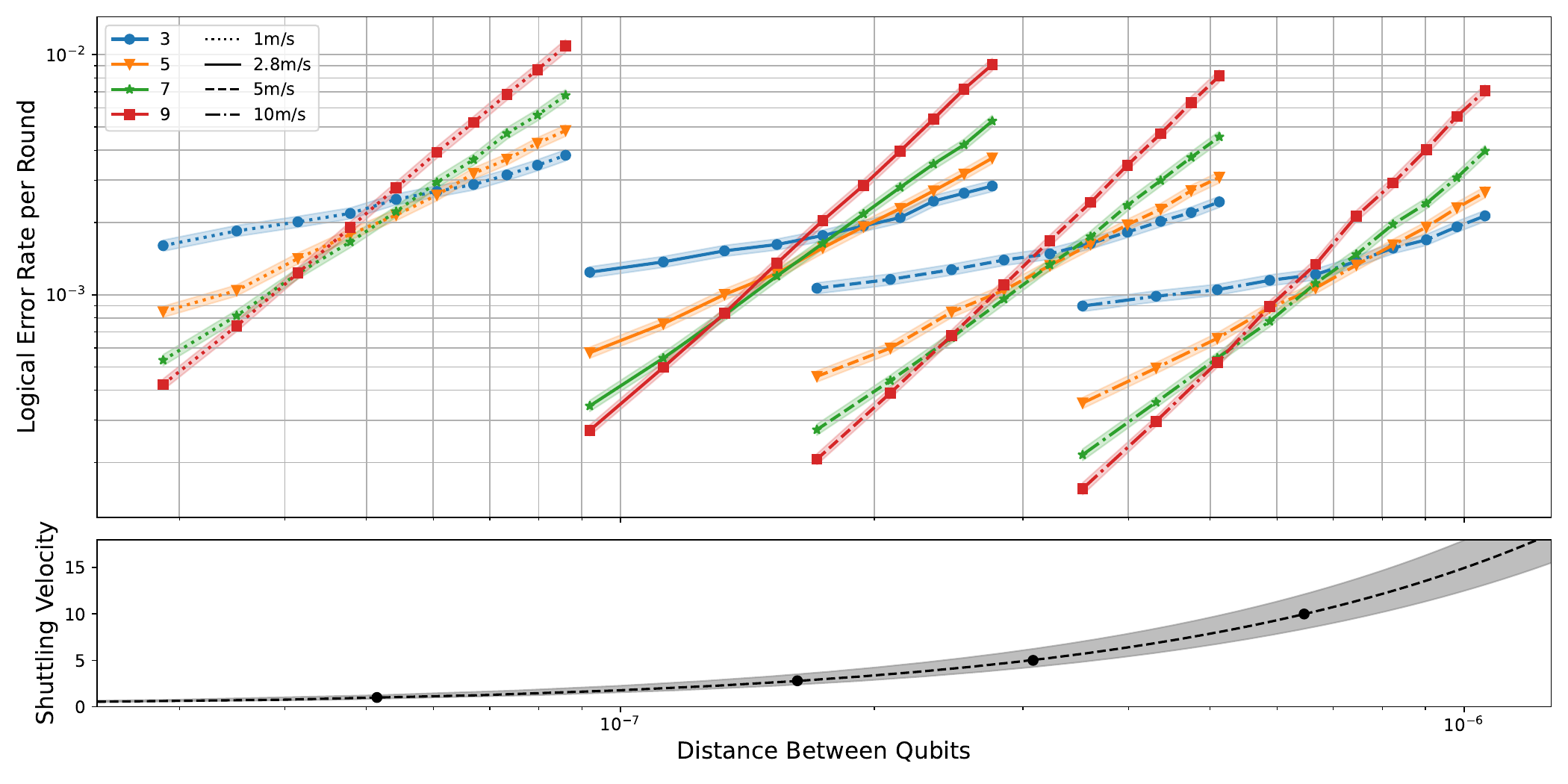}
    \caption{Logical error rate (top row) for four different shuttling velocities (1 -- 10 m/s) for several distance between qubits in the shuttling bus. The lower row represents the threshold (mean and standard deviation) for the shuttling velocity / distance between qubits tradeoff.}
    \label{fig:arch_parameters}
\end{figure*}

Figure \ref{fig:arch_parameters} depicts the logical error rate when exploring the distance between qubits ($d_{qu}$) for four different shuttling velocities, ranging from 1 to 10 m/s. The bottom row shows the threshold for each shuttling velocity computed as the mean among the intersections, with the shaded region representing the standard deviation.

As expected, a higher shuttling velocity leads to a higher tolerance for the distance between qubits. In the considered error model, there is no penalization for high shuttling velocities or low qubit distances. However, technology will limit the values that both parameters can take.

\begin{figure*}
    \centering
    \includegraphics[width=1\linewidth]{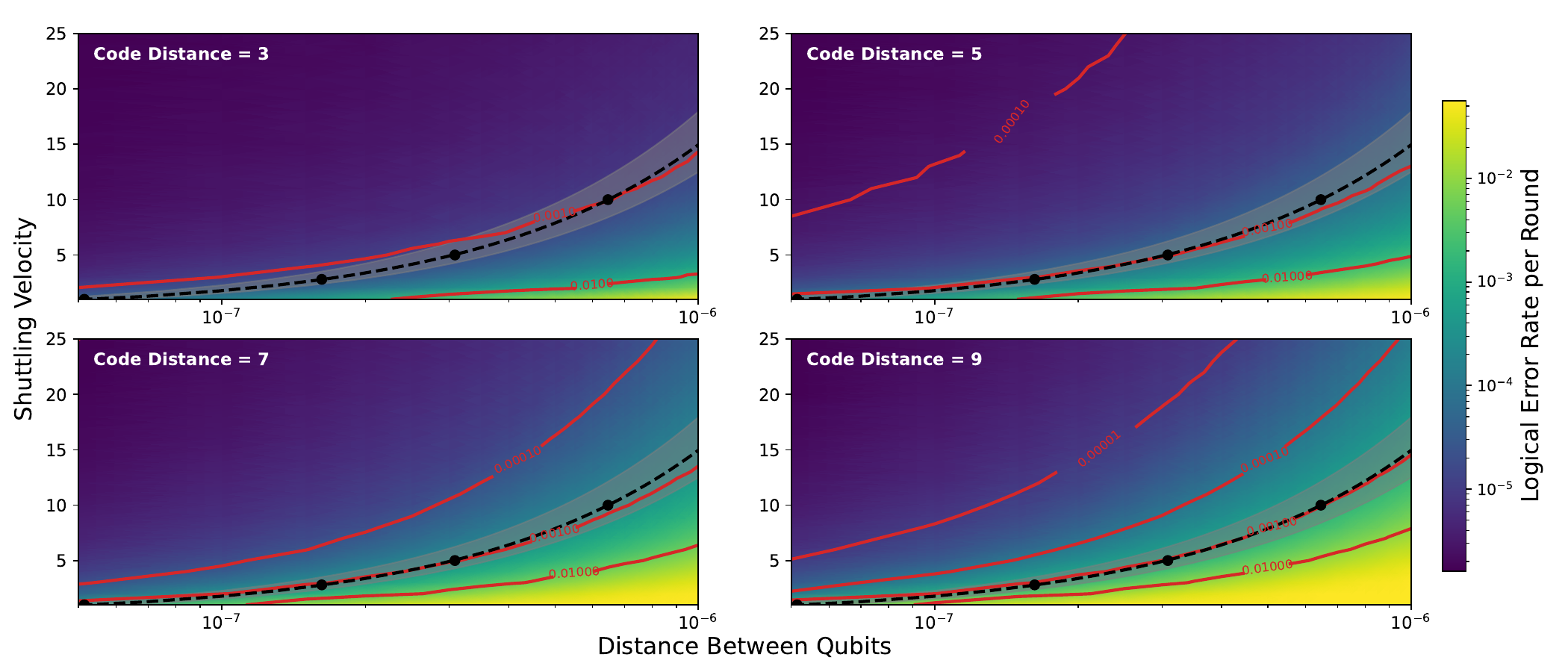}
    \caption{Logical error rate per correction round for distances 3, 5, 7, and 9 for different shuttling velocities, distance between qubits configurations. Red lines represent the isoclines for the logical error rate and the dashed black line depicts the threshold fit.}
    \label{fig:log_error}
\end{figure*}

To further explore the impact of the two considered architectural parameters, we explore the logical error rate per round obtained for a vast combination of shuttling velocities and qubit distances. Figure \ref{fig:log_error} shows the logical error rate per round for code distance 3, 5, 7, and 9 when exploring architectural parameters. The red and black lines depict the isoclines on logical error rate and the threshold fitted from the data in Figure \ref{fig:arch_parameters}, respectively.

As expected, when increasing the code distances, the logical error rate decreases for all the architecture configurations above the fitted threshold (dashed black line), and increases for the configurations below the fitted threshold.

\subsection{Suboptimal Architecture Synthesis}

After assessing, in the previous sections, the impact of the noise parameters and the architectural decisions on the performance of the surface code, in this section we explore the logical error rate for the three different architecture synthesis techniques discussed in Section \ref{sec:surface_code_synthesis}.

The noise parameters are set as in the previous experiments ($p=0.001$, ${T_2^*}_{QD} = 100 \mu s$, and ${T_2^*}_{bus} = 10 \mu s$), and we explore the three following architecture configurations:
\begin{itemize}
    \item $d_{qu} = 200nm$ and $v_{sh} = 5m/s$.
    \item $d_{qu} = 100nm$ and $v_{sh} = 10m/s$.
    \item $d_{qu} = 100nm$ and $v_{sh} = 20m/s$.
\end{itemize}

with the three proposed architecture synthesis techniques (\emph{Naive-topological}, \emph{Zig-Zag}, and \emph{Optimal}).

\begin{figure*}
    \centering
    \includegraphics[width=1\linewidth]{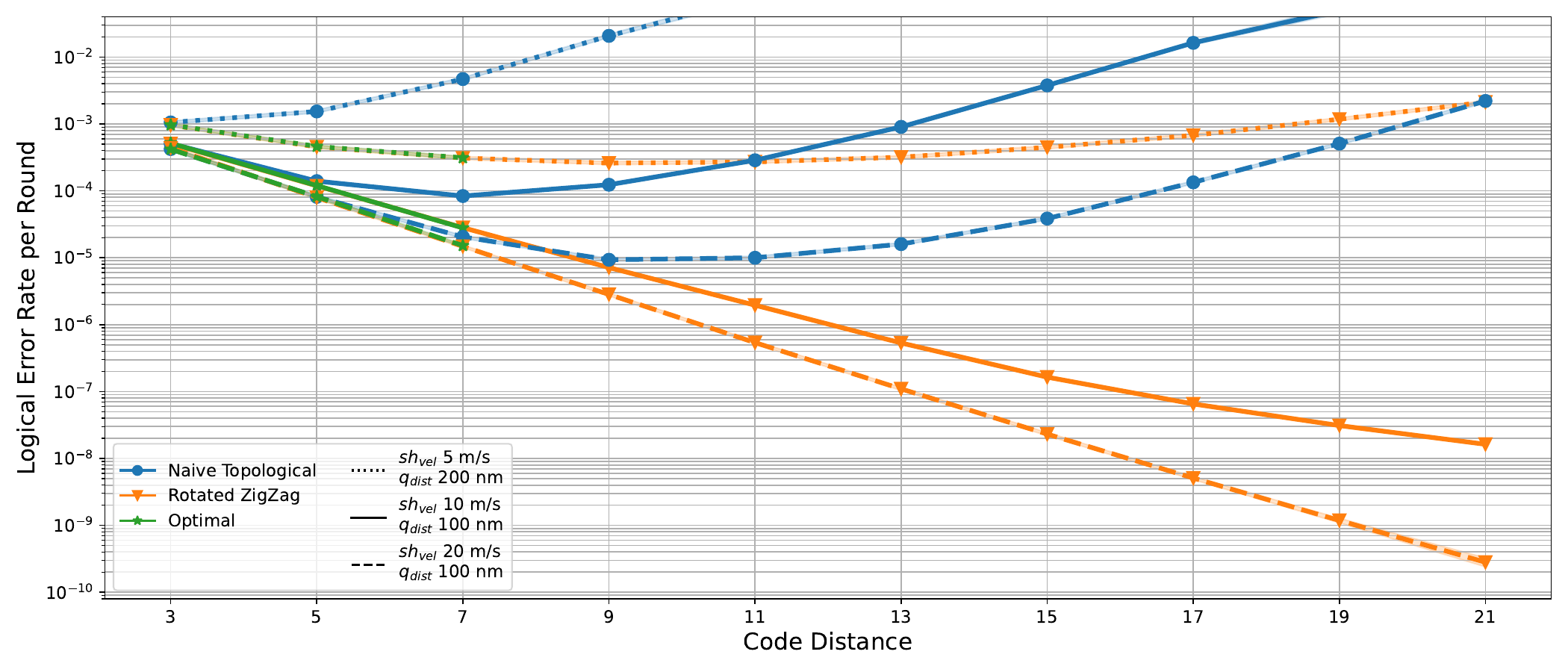}
    \caption{Logical error rate per round for the three proposed architecture synthesis techniques under three architecture configurations. Noise values are set to $p=0.001$, ${T_2^*}_{QD} = 100 \mu s$, and ${T_2^*}_{bus} = 10 \mu s$.}
    \label{fig:placement_impact}
\end{figure*}

Figure \ref{fig:placement_impact} shows the logical error rate per round when increasing the code distance for the three architecture configurations, and the three considered architecture synthesis techniques. Due to the computational cost of obtaining the \emph{Optimal} placement, we limit the exploration of such architecture up to code distance 7. For the \emph{Naive-topological} and \emph{Zig-Zag} placements we explore up to code distance 21, achieving logical error rates up to $2\times10^{-10}$.

The results obtained highlight the importance of synthesizing the optimal or close to optimal layout, since, for the same placement configuration and noise values, the \emph{Optimal} and \emph{Zig-Zag} placements obtain better logical error rates when increasing the code distance, while the \emph{Naive-topological} placement obtain worse logical error rates when increasing the code distance, meaning that the architecture configuration may be below the threshold.

With this exploration we conclude the analysis of the proposed architecture and highlight the under performance of the \emph{Naive-topological} layout synthesis, while showcasing the equal performance of the \emph{Optimal} and \emph{Zig-Zag} synthesis for systems up to code distance 7. Yet, we believe the \emph{Zig-Zag} placement matches the optimal one for larger code distances, but due to the computational cost of obtaining such optimal layouts this remains an unanswered question.

\section{Conclusions}
\label{sec:conclusions}

We introduced a complete methodology for synthesizing an optimal architecture for the rotated surface code on a one-dimensional spin-qubit shuttling bus. Our approach begins with a mixed-integer linear program (MILP) that precisely models the placement and shuttling constraints for syndrome extraction. This formulation yields exact solutions with relatively low execution time for code distances below $d=9$, allowing us to explore the structure of optimal encodings. Based on this structure, we proposed the \emph{Zig-Zag} heuristic for the placement of the rotated surface code, which scales linearly with the number of qubits and maintains the optimal cost in terms of total shuttling distance and cycle latency.

The MILP formulation at the core of the Quantum Reverse Mapping framework is not specific to the rotated surface code. The only requirement for applying our method to any quantum error-correcting code is that its syndrome-extraction interactions can be represented by a directed acyclic graph. Once this DAG is provided, the same optimization model directly applies, making the framework broadly general and adaptable to a wide range of error correction codes.

We assessed the three proposed architecture synthesis methods with detailed architectural metrics. Compared to a naive-topological placement, our methods reduce the total shuttling distance and round duration from quadratic to linear scaling with respect to the code distance. Additionally, our approach requires only two extra quantum-dot storage sites irrespective of the code size, whereas the naive method incurs spatial overhead linear to the code distance $d$.

To evaluate fault tolerance, we performed simulations using Stim under a noise model combining gate depolarization, quantum-dot dephasing, and shuttling-induced decoherence. We observed that, for realistic spin-qubit parameters, the synthesized architecture achieves logical error rates below $2\times 10^{-10}$ per round at code distance 21. These results demonstrate that spin-based quantum platforms are approaching the fidelity and coherence required for scalable fault-tolerant computation using surface codes.

This work underscores the importance of optimally encoding a single logical qubit before scaling up to large architectures. The reverse-mapping shuttling-bus architecture design provides a robust foundation upon which higher-level heuristic compilers can operate, ensuring that the scaling of quantum computers is supported by the strongest possible logical layer.

\begin{acknowledgments}
Authors acknowledge funding from the EC through HORIZON-EIC-2022-PATHFINDEROPEN-01-101099697 (QUADRATURE). 
EA acknowledges funding from the Spanish Ministry of Science, Innovation and Universities and European ERDF (QCOMM-CAT-Planes Complementarios), Generalitat de Catalunya, and the ICREA Academia Award 2024.
SA acknowledges funding from the EC through HORIZON-ERC-2021-101042080 (WINC). 
EB acknowledges funding from Science Foundation Ireland under Grant 21/RP-2TF/10019.
CGA acknowledges funding from the QuantERA grant EQUIP with the grant number PCI2022-133004, funded by Agencia Estatal de Investigación, Ministerio de Ciencia e Innovación, Gobierno de España, MCIN/AEI/10.13039/501100011033, and by the European Union NextGenerationEU/PRTR, and from the Spanish Ministry of Science, Innovation and Universities and European ERDF under grants PID2021-123627OB-C51 and PID2024-158682OB-C31.
PE acknowledges funding from the predoctoral FPI-UPC grant supported by UPC and Banco Santander.
\end{acknowledgments}

\section*{Data Availability}
All code developed and used in this work, including the optimization framework, architecture synthesis routines, and quantum error-correction simulations, is publicly available on Zenodo at \url{https://zenodo.org/records/17405430} \cite{escofet_2025_17405430}.

\appendix

\section{Applicability to Color Codes}
\label{sec:color_codes}
In this work we propose a methodology based on the formulation of a MILP optimization problem to synthesize an optimal architecture for the surface code in a shutting bus architecture. However it may be unclear which of the proposed methods can be applied to other QEC codes and which ones are intrinsic of the surface code.

The only claim only applicable to the surface code is that the \emph{Zig-Zag} synthesis method has an optimal performance (at least for code distances up to $d=7$). The other proposed architecture synthesis techniques (\textit{i.e.}, the \emph{Naive-topological} and the \emph{Optimal} one using the MILP formulation) can be directly applied to any QEC code for which we can construct the chain-preserving DAG (see Figure \ref{fig:dag} from the main text), which not only depend on the code, but also on the scheduling of the data-ancilla interactions.

To illustrate the generality of the MILP-based Quantum Reverse Mapping framework, we extend the approach to the family of 2D color codes \cite{bombin2006topological, landahl2011fault}. For these codes, we obtain both the lattice construction and the data–ancilla interaction ordering for the syndrome-extraction circuit from \cite{lee2025color}. This interaction structure naturally produces an acyclic dependency graph, fulfilling the only requirement needed to apply the MILP model.

The adaptation to color codes is therefore minimal: because their stabilizers have weight six, the syndrome-extraction circuit requires six interaction slices instead of the four used in the rotated surface code. Consequently, the number of time variables in the MILP increases from five to seven (six forward slices plus a return step), while all other constraints remain unchanged.

\begin{figure*}
    \centering
    \includegraphics[width=1\linewidth]{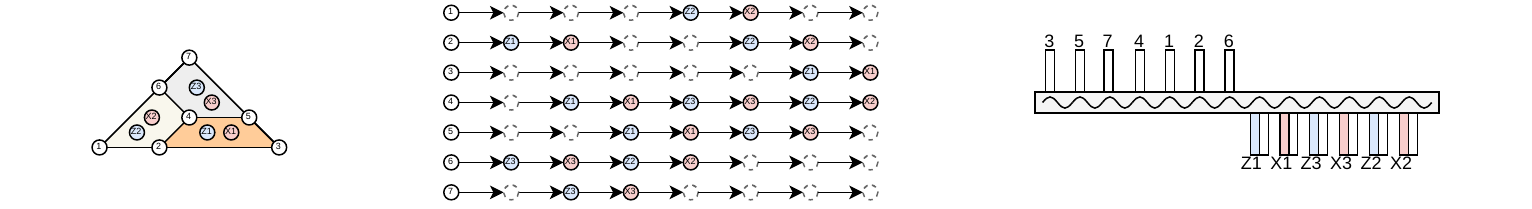}
    \caption{Stabilizer layout, ancilla–data interaction structure, and optimal architecture synthesis for the 2D color code of distance $d=3$. The six data-qubit interactions associated with each stabilizer define an acyclic dependency graph suitable for the MILP-based Quantum Reverse Mapping framework.}
    \label{fig:color_code_3}
\end{figure*}

\begin{figure*}
    \centering
    \includegraphics[width=1\linewidth]{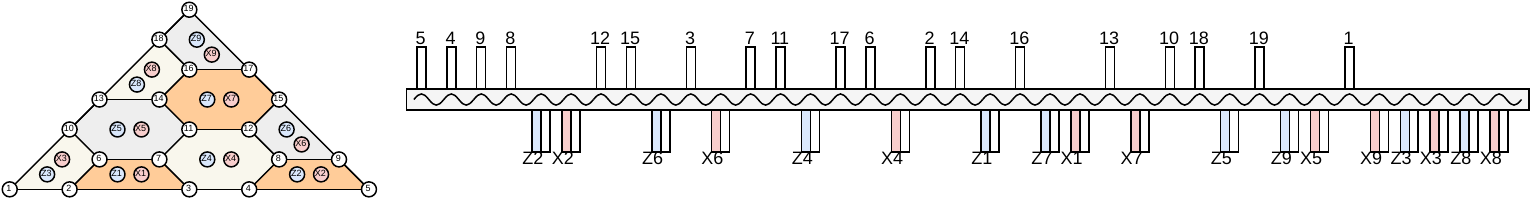}
    \caption{Stabilizer layout and optimal architecture synthesis for the 2D color code of distance $d=5$.}
    \label{fig:color_code_5}
\end{figure*}

\begin{figure*}
    \centering
    \includegraphics[width=1\linewidth]{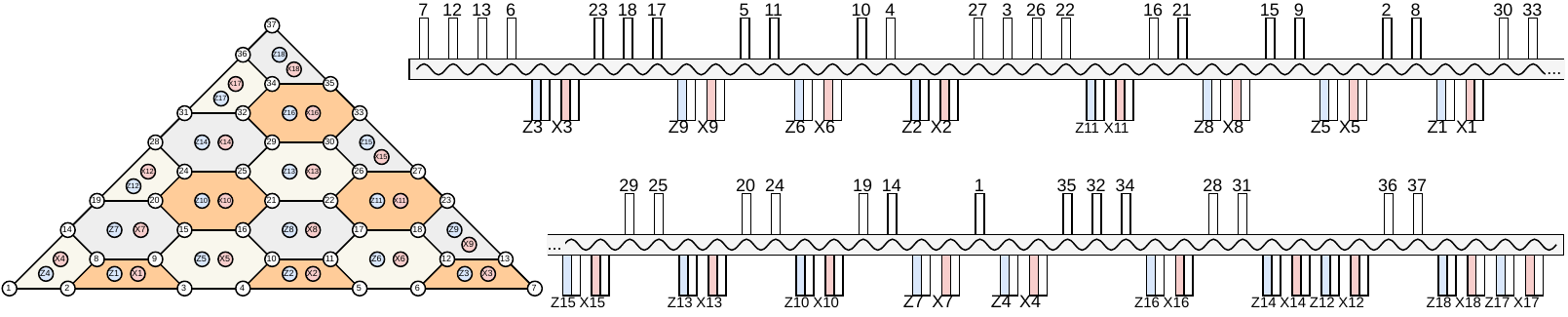}
    \caption{Stabilizer layout and optimal architecture synthesis for the 2D color code of distance $d=7$.}
    \label{fig:color_code_7}
\end{figure*}

Using this extended formulation, we synthesize optimal one-dimensional shuttling architectures for 2D color codes with distances $d=3$, $d=5$, and $d=7$. The resulting stabilizers and optimal placements are shown in Figures \ref{fig:color_code_3}, \ref{fig:color_code_5}, and \ref{fig:color_code_7}, respectively.

\bibliography{apssamp}

\end{document}